# Thriving in a Pandemic: Lessons Learned from a Resilient University Program Seen Through the CoI Lens

Zihui Ma[1*], Lingyao Li[2], John C.E. Johnson[3]

[1]Department of Civil and Environmental Engineering, University of Maryland, College Park, MD, zma88@umd.edu
[2]School of Information, University of Michigan, Ann Arbor, MI, lingyaol@umich.edu
[3]Department of Civil and Environmental Engineering, University of Maryland, College Park, MD, johnj85@umd.edu
[*]Corresponding author, zma88@umd.edu

**Abstract:** In March 2020, college campuses underwent a sudden transformation to online learning due to the COVID-19 outbreak. To understand the impact of COVID-19 on students' expectations, this study conducted a three-year survey from ten core courses within the Project Management Center for Excellence at the University of Maryland. The study involved two main steps: 1) a statistical analysis to evaluate students' expectations regarding "student," "class," "instructor," and "effort;" and 2) a lexical salience-valence analysis (LSVA) through the lens of the Community of Inquiry (CoI) framework to show the changes of students' expectations. The results revealed that students' overall evaluations maintained relatively consistent amid the COVID-19 teaching period. However, there were significant shifts of the student expectations toward Cognitive, Social and Teaching Presence course elements based on LSVA results. Also, clear differences emerged between undergraduates and graduates in their expectations and preferences in course design and delivery. These insights provide practical recommendations for course instructors in designing effective online courses.

**Keywords**: Distance education and online learning; Data science applications in education; Improving classroom teaching; Learning communities; Teaching/learning strategies

## 1. Introduction

The COVID-19 pandemic brought an unexpected and profound impact on the education sector. In March 2020, universities worldwide were compelled to shift to online learning as a result of the pandemic. By July 2020, The United Nations Educational, Scientific and Cultural Organization (UNESCO) reported that 111 countries had closed schools, affecting over 1.07 billion students, approximately 61% of the global student population (UNESCO, 2020). This abrupt transition forced students and educators to use online platforms like video conferences and discussion forums to facilitate their education.

Amid this shift, a global survey conducted by Pearson (*The Pearson Global Learner Survey*, 2020) illuminated multifaceted challenges faced by educational institutions. These included pedagogical adaptations, technological integrations, and the alignment of educational offerings with the swiftly evolving expectations of the student body (Fewella, 2023; Iglesias-Pradas et al., 2021; Reshi et al., 2023). Even after the pandemic, the widespread adoption of online learning continued to influence the global education landscape (*The Pearson Global Learner Survey*, 2020)， emphasizing the importance of aligning educational delivery methods with students' evolving needs and expectations.

One traditional approach to understanding students' expectation relies on the collection of numerical ratings through course surveys (Brinkerhoff & Koroghlanian, 2007; Harris et al., 2014; Horspool & Lange, 2012; Landrum et al., 2021; Paechter et al., 2010). While these surveys are valuable, they often concentrated on specific aspects of e-learning courses (Hamdan et al., 2021; Kang & Park, 2022), potentially missing the holistic nature of the student experience in the online learning environment.

Another branch in this field focuses on the analysis of students' textual feedback, typically employing two approaches: sentiment analysis (Dolianiti et al., 2019; Hew et al., 2020; Li, Johnson, et al., 2022; Melba Rosalind & Suguna, 2022) and thematic analysis (Braun & Clarke, 2006; Caskurlu et al., 2021; Henry, 2018; Meinck et al., 2022). Those methods offer a more in-depth comprehension of students' perceptions of their attended courses. However, few studies have explored the impacts of the COVID-19 pandemic on student expectations, leaving a demand for more comprehensive research in this domain. Moreover, existing studies often offer only a snapshot of the situation based on surveys conducted shortly after school closure.

This study seeks to fill existing research gaps by conducting an investigation into expectations of students enrolled in the Project Management Center for Excellence at the University of Maryland toward online learnings during COVID-19 pandemic. Accordingly, this paper focused on elucidating answers to two primary research questions:

1. What discernible differences exist in the expectations and preferences of undergraduate and graduate students concerning online learning when considering an extended study timeline?

2. How have students' expectations and preferences evolved through the lens of the Community of Inquiry (CoI) framework amid the COVID-19 pandemic?

In pursuit of these objectives, the sample data collection process was executed through the distribution of online questionnaires administered to students participating in a selection of ten courses. These courses included both undergraduate and graduate levels and spanned four academic years.

Firstly, we employed statistical analyses to assess the trends in overall course ratings between the pre-COVID-19 and post-COVID-19 periods. Subsequently, a lexical salience-valence analysis (LSVA) was applied to capture changes in sentiment by analyzing the recurring terms in students' feedback. Lastly, to offer a comprehensive perspective on the impact of the pandemic on students' perceptions of course quality and their areas of concern, we adopted the well-established CoI framework. By combining these research methodologies, we are allowed to delve deeper into both quantitative trends and nuanced qualitative shifts regarding the online learning experience, shedding light on the changes in social, cognitive, and teaching presences within the digital classroom environment. Ultimately, this study offers practical guidelines for addressing students' learning preferences, thereby contributing to the design of more effective courses in the future.

# 2. Literature Review

## 2.1. The impact of COVID-19 on learning experience

The advancement in technology has transformed the learning environment. Online education is growing at an incredible rate, possibly due to its convenience and flexibility that traditional in-person classes cannot match. Recent research has explored whether online learning is more productive than traditional teaching methods. Some studies have found that students obtained higher grades in online learning than in face-to-face instruction (Cavanaugh & Jacquemin, 2015; Ladyshewsky, 2004). Others conducted meta-analyses to distinguish differences between the two learning environments (Means et al., 2013; Pei & Wu, 2019; S. Lockman & R. Schirmer, 2020).

However, the existing body of literature on the transition from face-to-face teaching to online learning generally assumes that instructional changes are thoughtfully planned by instructors (Charles et al., 2020; Gelles et al., 2020; Johnson et al., 2020). This assumption, while appropriate for stable educational settings, faced a significant test when confronted with unforeseen challenges posed by the rapid onset of the COVID-19 pandemic. In particular, the sudden transition left little time for preparation, presenting challenges for pedagogical adaptation and technological integration. It also engendered a scenario wherein educators, irrespective of their prior experience or preparedness for online teaching, had to swiftly navigate the complexities of digital pedagogy. In essence, the pandemic exposed the fragility of traditional educational models when faced with rapid and unanticipated disruptions, necessitating an agile and adaptable response from the academic community.

Prior studies revealed students' negative experiences with online learning during the pandemic. For instance, Baltà-Salvador et al. (2021) found that most engineering students were dissatisfied with the quality of their online education. They also observed that the students' most-reported feelings included discouragement, boredom, confusion, and worry. In another study with dental students at Harvard University, Chen et al. (2021) showed that students perceived a decline in their academic performance during the transition to virtual learning, with increased burnout and decreased engagement.

The sense of belonging plays a critical role in shaping students' learning experiences (Zengilowski et al., 2023). The lack of face-to-face interaction with their cohorts and immediate feedback from instructors has made it more difficult for students to stay on track with their course-work and receive the support they need to succeed. This could further affect their motivation and engagement during COVID-19 (Chan et al., 2020). Gherghel et al. (2023) conducted a study involving over 1000 first-year students who enrolled in online classes during 2020 and 2021. By investigating emotional and behavioral engagement, the study confirmed that emergency remote learning had a significant impact on reducing social interaction and decreasing student engagement during the initial phase of the pandemic. Furthermore, the limited access to the internet or technologies served as an additional barrier hindering students' effectiveness in online classes (Baticulon et al., 2021; Jæger & Blaabæk, 2020). Consequently, this transition has demonstrated challenges for students and highlighted the necessary of reevaluating instructional strategies to suit current online learning mode.

On the other hand, graduate and undergraduate students may experience different impacts from COVID-19 due to the distinct nature of their academic journeys and expectations (Artino & Stephens, 2009; Salta et al., 2022). As stated in Farias Bezerra et al. (2021), about half of surveyed graduates expressed satisfaction with their e-learning performance, whereas a relatively smaller proportion of undergraduates reported poor or very poor performance. Soria et al. (2020) found out graduate students tend to exhibit more adaptive self-regulated learning profiles compared to undergraduates. Furthermore, the study revealed that 39% of undergraduates perceived unclear expectations for online learning from their instructors, in contrast to only 23% of graduate students. These disparities emphasize the need for educators to design targeted support systems and instructional strategies, particularly in the post-pandemic educational landscape.

### 2.2. The methods of student expectations analysis

It is important to ensure that online courses align with students' expectations. Student satisfaction and engagement are pivotal factors that contribute to the overall success of educational programs (De Santos-Berbel et al., 2022; Elshareif & Mohamed, 2021). In the environment of online educa-tion, where physical presence of instructors and peers is absent, meeting student expectations becomes even more critical for fostering a positive and effective learning experience (Balloo, 2018; Toufaily et al., 2018; Wu et al., 2006).

Traditionally, the evaluation of whether online courses align with student expectations is based on numerical ratings through course surveys (Brinkerhoff & Koroghlanian, 2007; Horspool & Lange, 2012; Landrum et al., 2021). Typical surveys, such as the Student Expectations of Online Learning Survey (SEOLS) developed by Harris et al. (2014), have been instrumental in gauging student expectations. SEOLS comprises 43 questions, with respondents using a 5-point Likert-type scale ranging from 1 (strongly disagree) to 5 (strongly agree) to rate their agreement with each item. The scores assigned to each scale were then aggregated to a total scale score. The outcomes of such experiential assessments have underscored the reliability and effectiveness of SEOLS in scrutinizing student expectations across seven distinct scales.

Similarly, Paechter et al. (2010) conducted a survey from 2196 students, utilizing a six-point scale questionnaire and organized into four sections. Blau et al. (2017) employed an instrumentation that utilized a 7-point Likert-type response scale, ranging from 1 (strongly disagree) to 7 (strongly agree). These surveys, while valuable, often distill the richness of student feedback into quantitative scores, offering a broad but somewhat limited perspective on course quality. While these ratings provide insights into overall satisfaction levels, they may not capture the nuanced details of the student experience. Moreover, these studies have predominantly focused on students' experiences with specific aspects of e-learning courses, such as their interactions with instructors, experiences with particular learning management systems, or certain course characteristics. However, they might overlook more generalized aspects of a course that students consider significant and preferential in the context of online learning.

Therefore, researchers have endeavored to mine textual feedback from students to gain a deeper understanding of the specific aspects that resonate with students' expectations and those that may need improvement. Several researchers have delved into sentiment analysis of students' feedback (Dolianiti et al., 2019; Li, Johnson, et al., 2022; Melba Rosalind & Suguna, 2022; Wen et al., 2014). For example, Baddam et al., (2019) accessed students' sentiment to understand what course characteristics could influence student evaluations. Hew et al. (2020) adopted sentiment analysis and hierarchical linear modelling to analyze the course features of 249 randomly sampled massive open online courses (MOOCs) and the perceptions of 6393 students regarding these MOOCs.

Another common method is to explore thematic on students' feedback to reflect their experiences with these adaptations (Braun & Clarke, 2006; Caskurlu et al., 2021; Henry, 2018). This thematic analysis can help researchers shape insights into the effective aspects as well as challenges and barriers in the online learning environment. For instance, Bourdeaux and Schoenack (2016) interviewed 22 adult students and uncovered themes explaining their expectations and experiences in online courses. The results revealed that students favored clarity, respect, and intention course design. Henry (2020) conducted in-depth online interviews with 43 students to discuss their expectations of online learning. They revealed six key themes to describe online students' expectations, including Motivation, Ability, Circumstances, Interaction, Curriculum, and Environment.

However, very few studies have thoroughly explored the impact of COVID-19 on online learning and its influence on student preferences. Moreover, these studies were primarily conducted shortly after the closure of schools and higher education institutions, leaving a gap in the literature that calls for the collection of larger datasets covering more extended study periods. For example, Aucejo et al. (2020) conducted a survey involving 1500 undergraduate students to gauge the causal effects of the pandemic, but their sample data exclusively pertained to students enrolled during the spring 2020 semester. Similarly, Cicha et al. (2021) focused their research on first-year students, exploring the primary factors influencing students' perceptions in online education. On the other hand, Syauqi et al. (2020) presented an overview of students' perceptions, albeit limited to those who had undertaken courses before May 2020.

### 2.3. Community of Inquiry (CoI) Framework

The theoretical framework of Community of Inquiry (CoI) was first developed by Garrison et al. (1999) for constructing and sustaining online educational experiences. According to the theory, an educational CoI is a grouping of individual learners who collaboratively engage in purposeful crucial discourse and reflection to build personal meaning and substantiate mutual understanding. The CoI framework defines a procedure of nurturing interests and promoting learning experience through the practice of three independent elements: social, cognitive, and teaching presence (Garrison et al., 2001). The CoI elements are defined as below.

- Social Presence is the capability of participants to recognize the community, communicate purposefully in a trustful setting, and develop interpersonal relationships by projecting their personalities (Garrison et al., 1999)

- Teaching Presence is the design, facilitation, and direction of cognitive and social processes to acknowledge personally meaningful and educationally valuable learning outcomes (Anderson et al., 2001).

- Cognitive Presence is the extent to which learners can build and confirm meaning through sustained reflection and discourse (Garrison et al., 2001).

Previous studies focusing on CoI applications mainly converge to two themes, (1) explore how the application of CoI theory can benefit educational practices, and (2) refine CoI framework configuration by exploring and validating additional factors. For the former, multiple studies have investigated the importance of CoI elements in educational settings and explored how those CoI elements can interact to improve learning satisfaction (Southam et al., 2022; Yandra et al., 2021). For example, Patwardhan et al. (2020) examined how CoI elements could contribute to student satisfaction in a remote learning environment. Their results showed a significant relationship between teaching presence and learning satisfaction. Similar to this conclusion, Purwandari et al. (2022) revealed that teaching presence has a higher contribution to cognitive presence than social presence in online engineering education and suggested that engineering educators should be equipped with pedagogical competence using the CoI framework. In another study, Li, Johnson, et al. (2022) discovered that teaching presence and cognitive presence were more or less important depending on the student intent and course content. That teaching presence dominated courses looking to impart knowledge, and that there was an intrinsic link between teaching and cognitive presence for courses focusing on skill development. However, social presence remains the relatively low impact in both cases for MOOCs.

For the latter, studies aim to appraise the CoI framework (Zulkanain et al., 2020) and explore factors to refine the framework configuration (Borup et al., 2022; Shea et al., 2012). For example, Kovanović et al. (2018) surveyed five MOOCs and proposed three additional factors, including course design, group affectivity, and resolution phase of inquiry learning, to teaching presence, social presence, and cognitive presence, respectively. Kaul et al. (2018) analyzed the discussion forum posts from one MOOC and suggested capturing more factors in distance learning for the CoI framework, such as the effects of course design and the engagement of learners. Most reviewed studies focus on refining the present CoI framework by proposing more nuanced factors to CoI elements, but a few attempts suggest adding new CoI elements and address the necessity when applied to the practical learning environment. For example, Cleveland-Innes and Campbell, (2012) recognized the development of new competencies due to the advancement of technologies. As a result, they thought emotional presence should be a fundamental element in the CoI framework. Wertz (2022) introduced learning presence as an additional element to the CoI framework by showing a strong relationship between learning presence and cognitive presence.

However, most of our reviewed studies discuss the applications of the CoI framework in the environment of distance learning (Kaul et al., 2018; Li, Johnson, et al., 2022) and blended learning (Gagnon et al., 2020; Liu & Deris, 2022). This is possibly because the CoI framework was initially designed for online learning experience. Studies applying the CoI framework to classroom settings are closer to the present study but less common in the literature. A typical study collected student feedback from 13 classes of a modern technology course and applied the refined CoI framework that included emotion presence (Zou & Zhang, 2022). This study found that the CoI theory was suitable for classroom settings where emotional presence was the most common, followed by cognitive and teaching presences, but social presence was the least common.

## 3. Material and Methods

### 3.1. Data Collection

We conducted an investigation at the Project Management Center for Excellence (PM Center) at the University of Maryland College Park from 2018 to 2021. The Project Management Center for Excellence had been teaching online since 2005 and was a pioneer in taking engineering education online through multiple delivery models. The faculty had been routinely improving course designs with optimal online-native teaching methods, such as flipped classrooms, modularized learning objectives, strong website and orientation, and active learning through projects. For many years, faculties from the Project Management Center at the University of Maryland avoided what is typically considered the standard pedagogical model of "sage-on-the-stage" models that emphasize in-person lectures with instructors speaking to an audience of students and where students are expected to read before and do assignments after the lecture. Instead, the faculties focused on flipped classrooms, discussion and workshops in-class, and applied learning through projects.

We focused on ten core courses (five undergraduate and five graduate) that were taught by the same instructor. We distributed an online questionnaire to both graduate and undergraduate students at the end of each semester to gather data on student course feedback. To ensure consistency in curriculum and instruction, only fall semesters were included for the subsequent analysis, as the 2020 spring semester was heavily affected by the COVID-19 pandemic. Based on the collection of survey responses, we defined two periods: (1) the first period covered fall semesters in 2018 and 2019, (2) the second period covered fall semesters in 2020 and 2021.

We distributed the survey to a total of 1253 students enrolled in the selected courses, in which 737 students were enrolled in Period 1 (T1), and 516 enrolled in Period 2 (T2). A required sample size of 577 students was calculated based on a 95% confidence level and a 3% margin of error, with 436 students and 348 students for each measurement. We received responses from 818 students, 465 in T1 and 353 in T2, meeting the intended sample size. The study ensured the protection of personal data by obtaining consent from all recipients and guaranteeing participant anonymity and data confidentiality. The collected data did not include any personal information of the students. Table 1 presents the distribution of the academic year, the number of student registrations, and responses for the undergraduate and graduate cohorts.

Table 1. The distribution of collected data information.

| Academic year | Group | Registered number | Responses number |
| --- | --- | --- | --- |
| Period 1(T1) | Graduate | 347 | 152 |
| | Undergraduate | 390 | 364 |
| Period 2 (T2) | Graduate | 256 | 114 |
| | Undergraduate | 209 | 239 |

### 3.2. Survey Development

We focused on three significant aspects: (1) student, (2) course, and (3) instructor, as presented in Table *2*. Questions in "Student Information" aim to investigate students' learning progress and abilities in the course. Questions in "Class Information" aim to understand how students feel satisfied with the class materials, such as assignments, exams, textbooks, and readings. Questions in "Instructor Information" aim to investigate the instructor's abilities in teaching materials for students. In addition, we added question 24 to examine students' weekly efforts in a course and questions 25, 26, and 27 to solicit students' constructive opinions regarding the course development.

Table 2. Survey design.

| Category | Survey question |
|---|---|
| Student Information | Q1: I completed all assignments such as homework, projects, and case studies on time. |
| | Q2: I finished reading all required readings before the mid-term and final exams. |
| | Q3: I read additional readings from the reading list and beyond the assigned material to further my knowledge in this subject area. |
| | Q4: I attended all classes. |
| | Q5: When I had questions or concerns, I raised them directly and immediately with the instructor. |
| | Q6: This class challenged my ability, such as thinking in new ways. |
| Class Information | Q7: Class was well organized with clear objective(s). |
| | Q8: Class presentations, activities, and assignments fulfilled the class objectives. |
| | Q9: Assignments such as homework, project and case studies helped with understanding of the material. |
| | Q10: Feedback on assignments such as homework, project and case studies was helpful for me to learn. |
| | Q11: Exams measured what I learned. |
| | Q12: Textbook and supplemental readings, etc. are appropriate for the course. |
| | Q13: Course added to my knowledge and abilities in this subject area. |
| | Q14: The learning from the course fits my expectation(s). |
| | Q15: I would recommend this course to others. |
| Instructor Information | Q16: Instructor presented the course material clearly with good examples and illustrations. |
| | Q17: Instructor stimulated interest in the subject. |
| | Q18: Instructor used class time effectively. |
| | Q19: Instructor challenged me to do my best. |
| | Q20: Instructor encouraged questions and class participation. |
| | Q21: Instructor clearly answered questions with good examples and illustrations. |
| | Q22: The pace of teaching was appropriate. |
| | Q23: Instructor was available for consultation outside of the classroom. |
| Effort | Q24: How many hours do you averagely spend each week for this course? |
| Open-ended questions | Q25: What did students like most about the course? |
| | Q26: What did students like least about the course? |
| | Q27: What additional constructive feedback did students offer? |

Questions 1 to 23 are 5-interval scale questions asking about satisfaction, including strongly disagree, disagree, neutral, agree, and strongly agree. Question 24 is a ratio scale question asking respondents to respond measurably. The last three questions are open-ended, aiming to gain insights from students regarding their opinions about the course. For subsequent quantitative analysis, we replaced strongly disagree, disagree, neutral, agree, and strongly agree with 1, 2, 3, 4, and 5, respectively.

### 3.3. Sentiment Classification

We selected a sentence as the unit for sentiment classification since a student review could cover multiple facets of CoI. We used sentence endings, including a semicolon (;), a period (.), an exclamation point (!), a question mark (?), and an ellipsis (…) to chunk each student review into multiple sentences. In this way, we could identify distinct CoI elements based on the predetermined word patterns in Table 3.

We followed an iterative development process as presented in Li, Johnson, et al. (2022) to collect word patterns. The process started with a brainstorming session where we applied a top-down estimation based on subject matter expertise to map keyword terms to CoI elements. Then, we proceeded with a bottom-up process by manually reviewing the top 1000 most frequently discussed words from students to ensure the coverage of keywords (excluding common and ambiguous words). During this process, we manually reviewed each of the top 1000 most occurring words and assigned related words to the CoI elements. Subsequently, we performed a final review of the word library to make alterations and additions to incorporate related words. This process encompasses a top-down and bottom-up approach, forming an iterative cycle, with keywords and CoI elements mapped for NLP analysis. As we progressed through three iterations involving developments and modifications, the word library was eventually finalized. Examples of CoI identification and sentiment classification are presented in Table 3. CoI elements extracted from the first row are Teaching presence; CoI elements from the second row are Cognitive presence; and CoI elements from the third row are Social presence.

Table 3. Representative examples of sentiment classification and CoI identifications.

| CoI elements | Keywords | Example | Sentiment |
|---|---|---|---|
| Teaching | class, material, professor, lecture, topic, concept, case, reading, video, instructor, content, teach, online, workload, course, schedule, zoom, slide, style, tool, background, example(s), feedback, study, session, inclass, schedule, speaker, lecturer, record, understand, guest | *"I believe that a bit more examples will help. Many of the concepts and formula dont have examples to which students can refer. Those examples only have final answers to it, not steps on how to reach it. If that can be provided, that will so much improve the understanding of the materials."* | Negative |
| | | *"Very good and clear explanations made by the instructor. Excellent reading materials and resources. Very funny and informative classes."* | Positive |
| Cognitive | time, learn, assignment, work, question, study, quiz, homework, knowledge, grade, exam, test, midterm, project | "*Get rid of yellowdigg. Don't have a homework assignment due the same day as a case study. Perhaps try to compact the lectures and get more to the point. Also, the midterm was very messed up and a whole portion wasn't even applicable to the class. Perhaps this can be completely revamped and maybe even the final if it turns out the same*" | Negative |
| | | "*The case study based learning is something different. Also, there are no* <u>*midterm*</u> *and a final* <u>*exam*</u> *which really ease a lot of pressure and helps me focus on learning the concepts and not performing*" | Positive |
| Social | student, group, team, discussion, conference, presentation, people, skill, job, life, career, classmate, session, activity, person, team, company, speech, participation, communication | "*Not easy to apply to the real* <u>*life*</u>*. For example, I can't ask my colleagues to do more than 10 tests so that I can "Know Others" better.*" | Negative |
| | | "*I really enjoyed the workshops with* <u>*classmates*</u> *as well as the project* <u>*teams*</u>*. It was a good representative of work models in project management.*" | Positive |

Next, we leveraged a sentiment tool called Valence Aware Dictionary and Sentiment Reasoner (Vader) (Hutto & Gilbert, 2014) to calculate the sentiment for a CoI element. Vader is an open-sourced and lexicon-based sentiment tool implemented under the MIT license, which can help decipher the emotion of a conversation by classifying it into a positive, negative, or neutral sentiment. Vader returns a weighted and normalized compound score between -1 (i.e., most extreme negative) and 1 (i.e., most extreme positive), which is calculated by summing the sentiment scores of each word in a sentence and then normalized to the range between -1 and 1 (Hutto & Gilbert, 2014). According to its documentation, suggestive threshold values for research to categorize a sentence are (Hutto & Gilbert, 2014):

$$sentiment = \begin{cases} positive & sentiment \in [0.05, 1.00] \\ neutral & sentiment \in (-0.05, 0.05) \\ negative & sentiment \in [-1.00, -0.05] \end{cases} \quad \text{Eq.1}$$

### 3.4. Lexical salience-valence Analysis (LSVA)

With identified CoI elements and their associated sentiments from each review, we used an approach called lexical salience-valence Analysis (LSVA) (Taecharungroj & Mathayomchan, 2019) to explore the underlying relationship between words in reviews and their impacts on CoI elements. The LSVA approach was first formulated by Taecharungroj and Mathayomchan (2019) to identify positive and negative words and impacts on sentiment in tourist attractions based on TripAdvisor reviews. Then, this concept was applied to examine customer experience on restaurant attributes (Mathayomchan & Taecharungroj, 2020) and airport quality service (Li, Mao, et al., 2022) using Google Maps reviews.

The LSVA approach analyzes the relationships between words and sentiments through the definitions of salience and valence. In addition to showing the frequency of words in positive or negative reviews, the LSVA approach can help assess the frequency of words in the corpus of documents and their effects on the overall sentiment (Taecharungroj & Mathayomchan, 2019) of each CoI element. The LSVA represents the salience and valence of a word as below,

$$Salience|_{word_i} = log_{10}\,(N_{total})|_{word_i} \quad \text{Eq.2}$$

$$Valence|_{word_i} = \frac{N(positive) - N(negative)}{N_{total}}|_{word_i} \quad \text{Eq.3}$$

where,

$N_{total}$ denotes the total number of student reviews where $word_i$ appears.

$N(positive)$ denotes the number of positive student reviews where $word_i$ appears.

$N(negative)$ denotes the number of negative student reviews where $word_i$ appears.

The LSVA approach calculates the salience of a word as the base-10 logarithm of its frequency and the valence as the division of the difference between $N(positive)$ and $N(negative)$ and total count $N_{total}$. In particular, the valence measures the degree of sentiment that a particular word appears in a corpus (Taecharungroj & Mathayomchan, 2019). Reviews that contain words with greater valence are more likely to be positive than those with negative words.

## 4. Results

This section is divided into two parts. In section 4.1, we address the first research question through a statistical analysis of overall course evaluations spanning the study period. This reveals trends in course dynamics between the study groups. In section 4.2, we delve into the second research question by exploring students' perceptions, with a focus on the CoI lens, using the LSVA analysis. This approach helps us understand how student experiences and preferences have evolved within the CoI framework amid the COVID-19 pandemic.

### 4.1. Statistical Analysis

To visualize the distribution of overall course ratings across both the undergraduate and graduate groups, we averaged the total ratings of each category for each course from all responders. This average, referred to as the category score, was then used to construct the boxplot illustrated in Figure 1(a) and Figure 1(b). It is import to note that outliers beyond two standard deviations from the mean were removed before the calculation to ensure the statistical analysis remains robust and reflective of the majority of the data. Subsequently, we calculated the percentage change in ratings to understand the impact of COVID-19 on the course evaluation. Therefore, we calculated the category score $x_{category}$, where category = class, instructor, student, effort, alongside the semester variable $i$ to represent the year ($i$ = 2018, 2019, 2020, and 2021). The percentage change was calculated using Equation 4, and the corresponding results were shown in Figure 1(c) and Figure 1(d).

$$\%\, change_{course} = Avg[\frac{x_{category,semester\ i+1} - x_{category,semester\ i}}{x_{category,semester\ i+1}}] \quad \text{Eq. 4}$$

Overall, the findings reveal a consistency in ratings across the “class,” “instructor,” and “student” categories for both graduate and undergraduate courses. This observation is supported by a slight change observed within each category (Figure 1(a) and Figure 1(b))and a change of less than 5% (Figure 1(c) and Figure 1(d)). Based on Figure 1, we further observed that there is an increase in the median for undergraduate ratings in these three categories. This upward shift suggests an enhancement in half of the undergraduate courses.

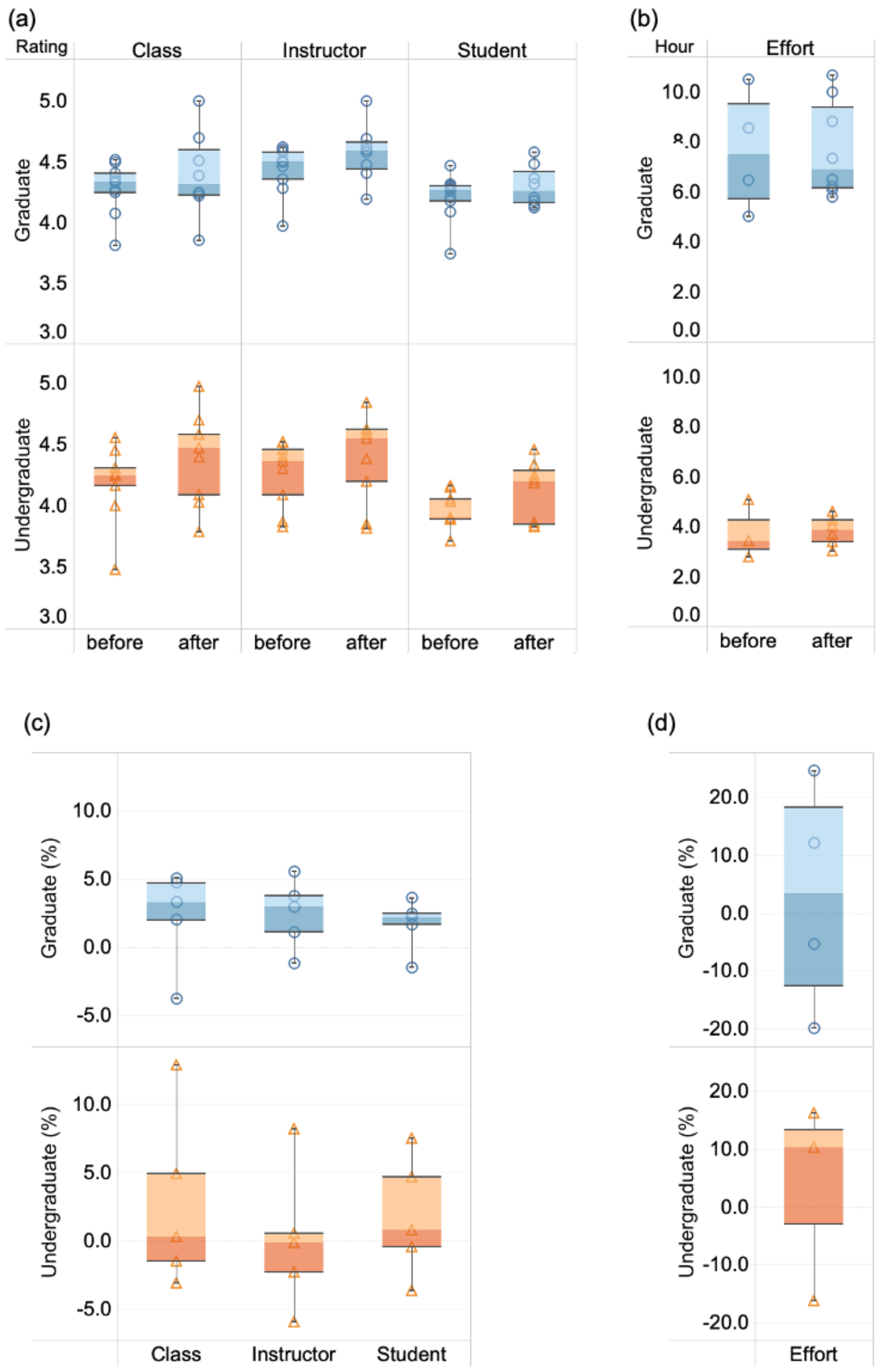

Figure 1. Average courses' rating score changed during COVID-19 for (a) "class," "instructor," "student" categories, and (b) "effort" category. Average percentages of courses' rating difference for (c) "class," "instructor," "student" categories, and (d) "effort" category.

Conversely, among the graduate courses, over half of courses shown a discernible improvement solely within the "instructor" category, indicated by the increased median average ratings.

Nevertheless, the upper quartile line in Figure 1(a) locates above its undergraduate counterparts, implying a higher level of satisfaction among graduate students. This divergence is also evident Figure 1(d), where all the percentage change of medians is positive in the graduate group while remaining close to 0 in the undergraduate group.

In terms of the "effort" category, it's intriguing to note that graduate courses consistently demand more effort, both before and after COVID-19, compared to undergraduate courses. For instance, as depicted in Figure 1(b), the graduate group reported a maximum of 11 hours per course, while undergraduates reported a maximum of 5.5 hours. Interestingly, despite the increased effort required for graduate courses, a shift is observed after COVID-19. Graduates noted a decrease in required effort, whereas approximately half of the undergraduate courses actually reported an increase. This is visually depicted by a median percentage change of nearly 10% in Figure 1(d).

Moreover, there is a notable increase in the variation of ratings after the onset of COVID-19. This is discernible from the wider dispersion of boxes within the "after" group when compared to the "before" group, as depicted in Figure 1(a) and Figure 1(b). This trend is excepted within the "effort" category. The broadening spread of the boxes implies that both graduate and undergraduate courses encountered more pronounced fluctuations in their design as the pandemic unfolded. In particular, Figure 1(c) and Figure 1(d) illustrates that undergraduate courses witnessed increases exceeding 10% as well as decreases of up to negative 5%. Despite this small but noticeable variability, across all the median lines of the left box remained within the confines of the comparison box plot in Figure 1(a) and Figure 1(b).

### 4.2. Course Feedback Analysis

This section aims to explore the evolving nature of students' needs and expectations within the study period from their feedback through the lens of the CoI framework. Using the LSVA approach, Figure 2 illustrates the relative importance of words related to the three CoI elements across two groups. Using Eq 2 and Eq 3, a higher salience score indicates that sub-topic words were more frequently mentioned in the dataset, while a higher valence score implies that a word was more positive based on students' feedback. To use the CoI framework, we calculated the difference in scores between the two study periods and presented the results in Figure 3.

(a)

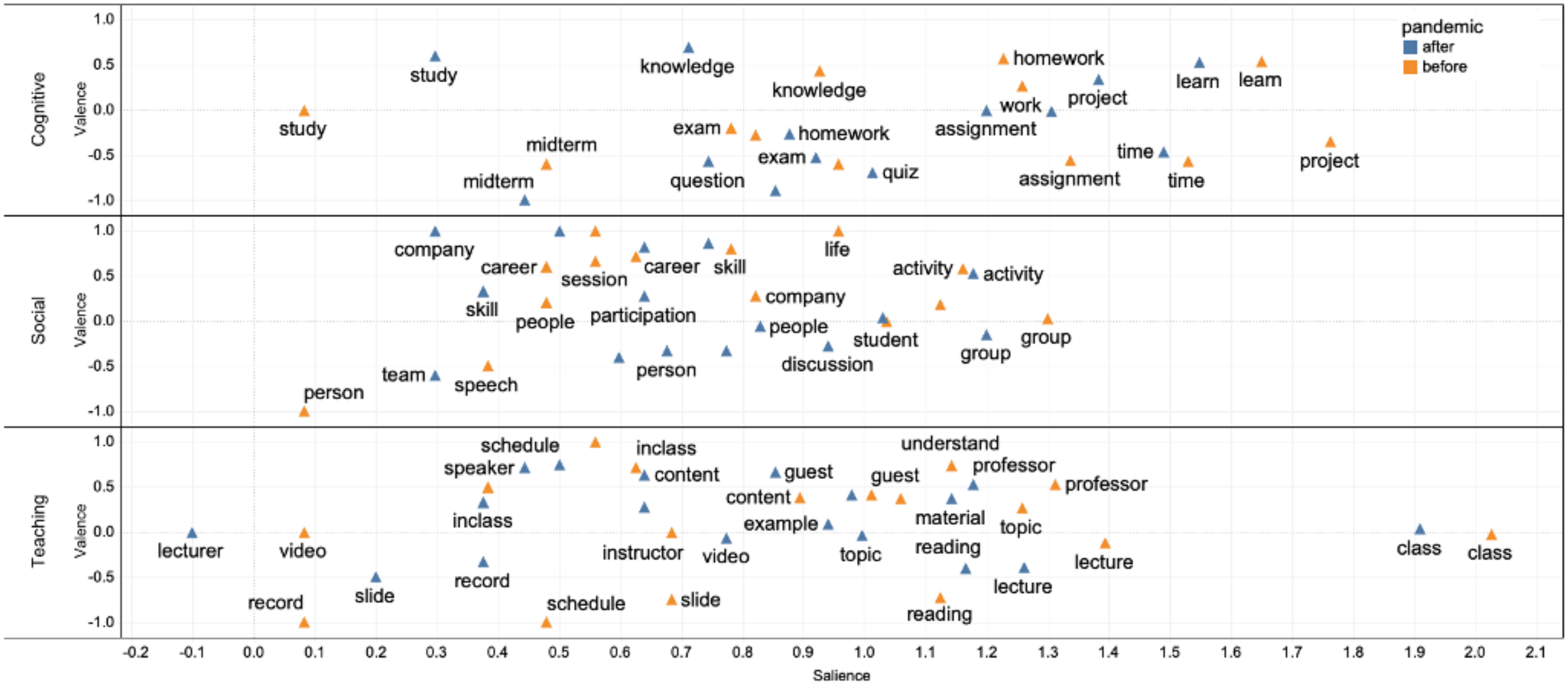

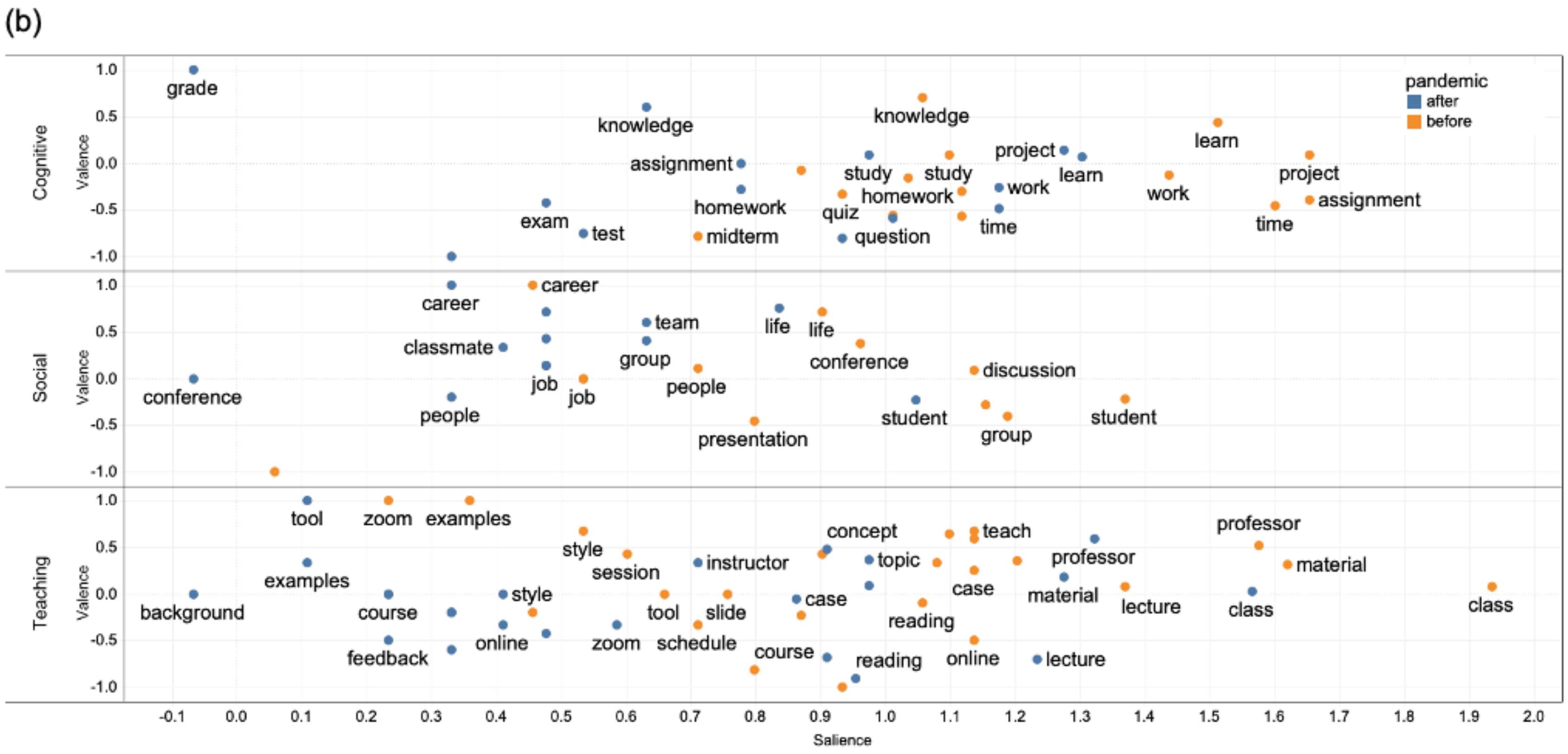

Figure 2. LSVA results for (a) undergraduate feedback, (b) graduate feedback.

(a)

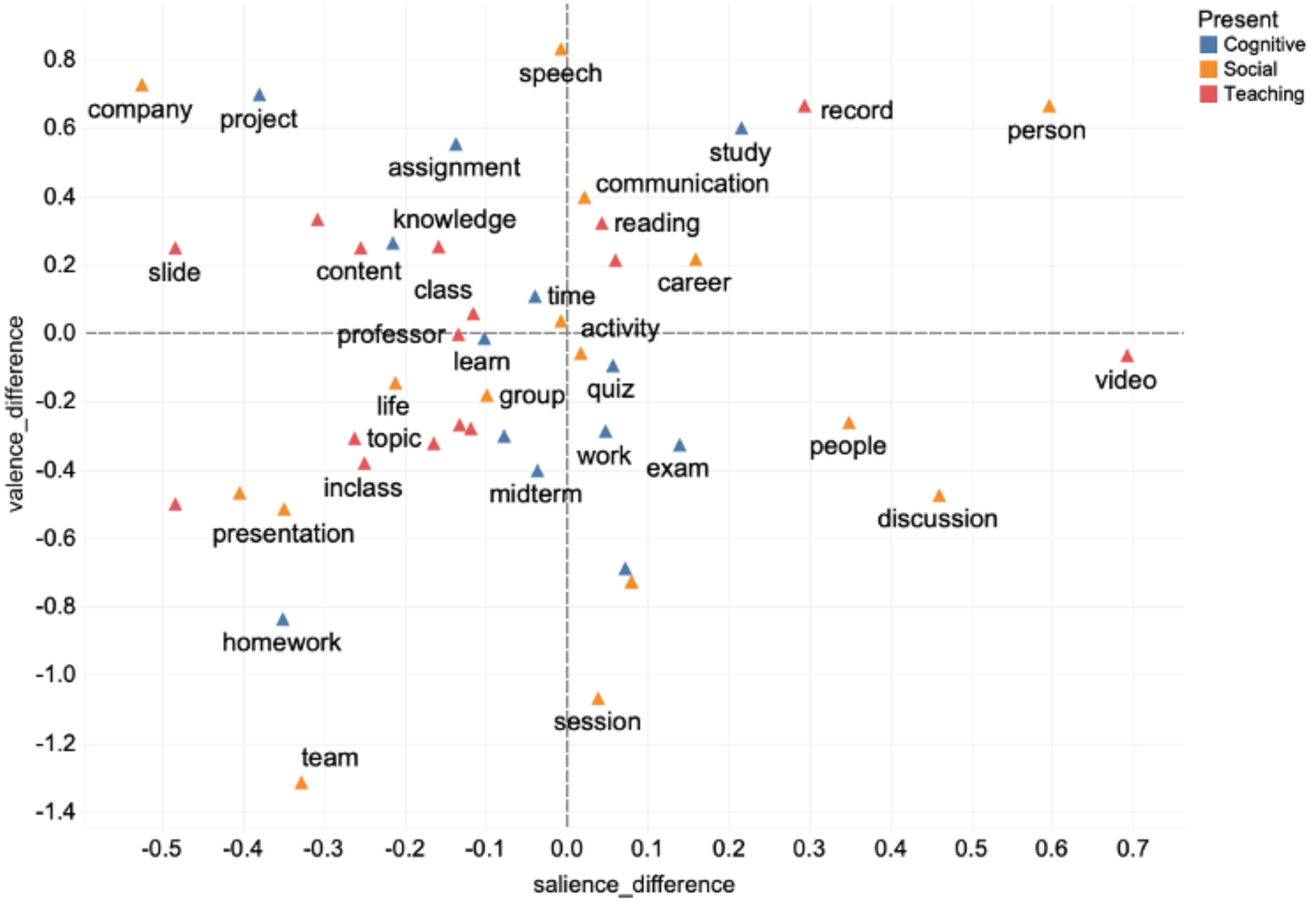

4

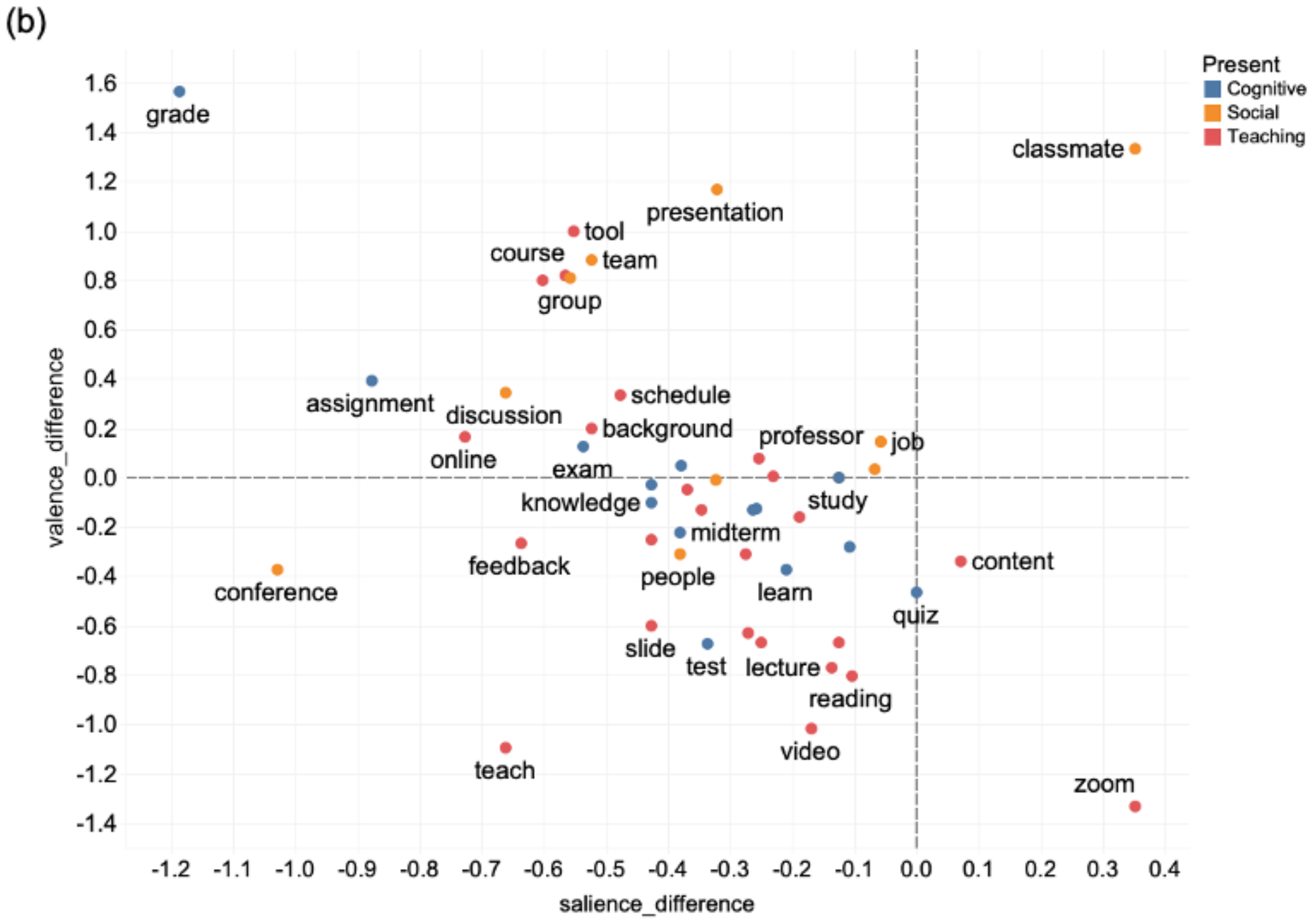

Figure 3. The difference of LSVA results for (a) undergraduate feedback, (b) graduate feedback.

#### 4.2.1. Cognitive presence

In the undergraduate group, negative valences were observed in most of the presented terms indicating a lack of preference for cognitive presence design elements like exams, projects, and homework during COVID-19 (Figure 2(a)) and most of them located in the fourth quadrant with declined valences in T2 (Figure 3(a)). For instance, terms such as "work," "homework," "exam," "question," "quiz," and "grade" experienced significant drops in valence, despite their frequent mention. Conversely, the words "assignment" "study," and "project" show higher valences implying positive changes after the pandemic.

Likewise, a negative trend in valence for the majority of terms indicates a declined in preference for cognitive presence course elements for the graduate group after COVID-19 outbreak. As shown in Figure 2(b), online education appears to have eased their time concerns with overloaded schedules at home, with a decrease in salience for the word "time" while valence remained almost the same. Additionally, graduate students seemed less interested in their grades in the T2 period. Although the term "grade" is highly positive, it has shown the most significant decline in salience (Figure 3(b)), meaning it wasn't being emphasized in class. However, graduates have the similar sentiment regarding the terms of "midterm," "quiz," "test," and "work" as their overall valence became negative during COVID-19, a need that stresses of COVID 19 caused students to no longer appreciate the challenge of these assignments. Observations with favorable terms like "knowledge" and "learn," related to critical thinking and learning skills, indicate that students wanted a learning experience emphasizing formative assessments and not one that was focused on evaluating their work.

#### 4.2.2. Social presence

The LSVA analysis conducted on undergraduate students revealed noticeable enhancements in terms of "activity," "student," and "person," as evidenced by the higher salience and increased valence (Figure 2(a)). Contrasting this positive trend, negative and decreased valence (Figure 2 (a) and Figure 3(a)) associated with terms such as "presentation," "discussion," "group," and "team" indicate that instructors might need to explore alternative means of coordinating group work to foster social presence. Furthermore, the drops in valence (Figure 3(a)) for the term "participation" suggests that when participation was emphasized in undergraduate courses it was a poorer experience or form of participation and not what students preferred.

On the other hand, it appears that graduate students were happier than undergraduates in terms of the social presence and class interactions, as indicated by an overall increase in scores for related terms. Figure 2(b) unveils a significant positive change in the sentiment of terms such as "classmate," "discussion," "team," "group" and "presentation," suggesting that graduates are enjoying the new communication methods, with the exception of video "conference" (which showed a decreased sentiment). Moreover, graduates felt there was a better alignment on how the course can benefit their long-term personal lives. This is reflected in the significantly higher valence scores for the terms "job," "life," and "career" compared to the pre-COVID-19 period (T1), but a decrease in salience. The only term that significantly increased in salience and valence (importance and positive preference) was "classmate." Instructors clearly emphasized social connections between classmates, and creating space for those interactions was highly valued by graduate students. Interestingly, feedback on the topic of "people" has flipped to negative sentiment were observed in Figure 3(b) which on further investigation revealed these comments tended to be less interest in connecting with people outside the classroom.

4.2.3. Teaching presence

Drawing insights from the LSVA outcomes concerning the undergraduate cohort, which are presented in Figure 2(a), teaching presence can be interpreted in terms of "topic," "example," and "lecture," which are strongly correlated with course components. The valence associated with these terms shows a significant negative difference between the two study periods (Figure 3(a)). The teaching presence can also be reflected by words related to course resources, such as "material," "slide," and "reading." While "slide" and "reading" still show negative sentiment, their valence has shown improvement. Terms tied to the teacher, such as "instructor" and "professor," remain nearly identical.

According to Figure 2(b), the overall teaching presence for the graduate group declined in the context of COVID-19 as indicated by the majority of words in red are located below the zero line for valence. In particular, the valence of the terms "material," "lecture," "case," "reading," and "slide" remarkably decreased. Compared to undergraduate students, graduate students appear to be less satisfied with online learning via “zoom” or “video,” as negative sentiments (Figure 3(b)) were observed in T2. While the term “professor” retained its relative valence, the terms “instructor,” “content,” and “style” were negatively affected by this sudden transition. Despite these passive changes, it is noteworthy that the terms “workload” and “schedule” received an increase in valence. This suggests that graduate course-related workload became more manageable.

## 5. Discussion

This study aims to examine the shift of student expectations and preferences regarding online and blended learning during COVID-19 pandemic. We collected 818 students’ responses who enrolled in the Project Management program at the University of Maryland over a four-year period and explored two research questions.

The first research question explored the overall ratings from undergraduate and graduate students. In terms of the three categories (class, instructor, student), our results revealed no significant differences, which suggests students’ felt that the PM Program courses adapted well during the pandemic. In the “effort” category, graduate courses commonly required more time than undergraduate courses, and workloads general went up (for undergraduate) or nearly stayed the same (for graduate) during the pandemic in T2. This means that the courses maintained the rigor and adapted well to COVID-19, allowing for a meaningful dive into the shifts in preferences by students during COVID-19 based on this sample group.

However, there was an increase in the variation of course ratings during T2. For instance, while the highest undergraduate rating in the “class” category experienced a positive change of around 10%, the lowest saw a decrease of approximately 4%. This implies that variations in course adaptation during COVID-19 in T2 were likely influenced by various factors, such as changes in course design, teaching performance, or other aspects related to the learning environment.

The second research question dove into the changes in student perspectives throughout the study period by employing the CoI framework. Our analysis confirmed that the overall student satisfaction of the Project Management Center for Excellence remained effective during the pandemic and thereafter. This positive trend was evident in the predominantly positive valence scores associated with most sub-topic words.

Digging deeper, our analysis also showed differences in student preference across various CoI elements. Specifically, our study revealed a significant decline in preference for cognitive presence-related course design elements among both undergraduate and graduate students following the pandemic. Undergraduate students placed greater emphasis on course outcomes, as evidenced by the higher salience of the term “grade”. Moreover, the findings indicate that students may have a preference for assignments or projects that have a formative nature over quizzes or exams that are summative in nature. There are likely two reasons. First, these activities may stimulate greater engagement among students and foster interaction with both their peers and instructors, educators may consider incorporating more interactive course activities to enhance cognitive presence for undergraduate students and increase social interactions (Social presence went up in valence significantly). Secondly, the students are more interested in applying and doing what aligns with their careers than being evaluated and likely appreciated the relaxation in grading policies. Interestingly, graduate students appeared to place less emphasis on “grades” but demonstrated the same pattern where valence for formative assignments like projects that remained the same or improved, while summative assignments like midterms and exams that test materials decreased significantly in student preference (e.g., valence).

In terms of social presence, graduate students exhibited relatively positive feedback compared to undergraduates, as reflected in the overall higher scores for terms related to social interaction. Undergraduates displayed a negative sentiment concerning terms, such as “presentation,” “discussion,” “group,” and “team,” suggesting that their expectations regarding collaboration activities were not met. This finding suggests that the medium in T2 during COVID-19 reduced the quality of these interactions. However, there was significant increase in elements relating to “career,” “communication,” and “speech” suggesting an emphasis on alignment of the materials with the students’ professional goals. While graduate students were content with the communication methods during the pandemic, focusing heavily on connections with classmates in a very positive way, and they expressed a similar interest in how the course can benefit their professional goals as well.

Both undergraduate and graduate students demonstrated a significant preference increase for most social elements, and in particular those elements that connect in-classroom learning to that which can help them in their careers. In graduate courses, this means connecting with peers who can share insights and support while tackling challenges at work and home during COVID-19. For undergraduates, this means preparing for their first big career steps.

Teaching presence for undergraduate students showed a decrease in the selection as evidenced by a decrease in sentiment for “topic,” “example,” and “lecture.” Meanwhile, satisfaction with the ‘instructor’ remained constant. This suggests that the medium for communicating these topics was not as effective, and that perhaps some topics and examples were not as meaningful during the COVID-19 time period.

In addition, our results demonstrated that graduate students weren’t as satisfied with the course materials evidenced by a remarkable decrease in terms such as “material,” “case,” “reading,” and “slide.” This stood out as important to students, suggesting that either students were not as willing or happy to engage with readings and case studies in the COVID-19 time period. However, the flexibility to do the work was clearly appreciated in T2 during COVID-19, as indicated by the presence of valence words such as “workload” and “schedule” above the zero line, students appear to be content with the current course commitment, possibly due to reduced time pressure.

These shifts in emphasis an acceleration of transformation in the post-COVID classroom, aligned with several global surveys conducted by institutions such Pearson (*The Pearson Global Learner Survey*, 2020), UNESCO (UNESCO, 2020) and UNICEF (UNICEF, 2021). These findings suggest that traditional teaching methods need to be radically restructured to align with the evolving needs and expectations of students. Crucially, this transformation is not a one-size-fits-all solution. Undergraduates lean towards practical application, aiming to become “job ready,” while graduate students prioritize networking and job support, recognizing the importance of social connections in their career paths. Both groups share a common desire for education that prioritizes their personal development over rigorous evaluation and demands flexibility in acknowledgment of their demanding lives.

### 5.1. Implications for educators

This study provides valuable insights into students’ perspectives on online learning amid COVID-19. It also offers practical recommendations for educators in designing effective online courses from the perspective of the CoI framework. First, educators can utilize the proposed survey instrument to assess course performance from students. The questionnaires cover four categories that capture the key elements of students' expectations in course learning. Additionally, this instrument enables educators to delve into broader facets of student responses regarding their expectations and preferences in the online learning experience. This approach is particularly beneficial for educators to evaluate online learning in response to emergency events, given that our analysis was conducted among students who have experienced the COVID-19 pandemic.

Second, the educator should address the evolving expectations of students. Social presence in the educational context holds distinct significance for different student groups. For undergraduates, it revolves around that pivotal moment of securing their first job, making the connection between academic learning and practical application more tangible than ever. To enhance their educational journey, it's advisable to incorporate more hands-on experiences and real-world case studies into the curriculum. Graduates, on the other hand, seek to harness the power of social presence for networking and mutual support among their peers in the professional world, underlining the paramount importance of social connections in career development. Therefore, we recommend that creating opportunities for networking and peer support in professional settings, such as facilitate platforms for alumni connections, mentorship programs, and industry-related events, to help students build social present. Regardless of their academic level, providing flexibility in schedules and workloads is essential to allow students to adapt education to their increasingly busy lives, ensuring a positive learning experience.

Cognitive presence intertwines with social presence. Students exhibit a notable preference for formative exercises, such as projects and assignments, emphasizing their preference for "learning by doing" rather than passive study of course materials. This preference aligns with their overarching goal of becoming "job ready" as they recognize that practical experiences provide a richer learning environment than mere theoretical understanding. Summative assessments, conversely, elicit a sense of disfavor among students, highlighting a desire for a more practical and applied approach to their education. In light of these insights, we recommend that undergraduate educators empower students to lead or co-lead course activities to enhance cognitive presence. For graduate educators, incorporating additional formative course resources is advisable. In both scenarios, fostering an environment that supports both social and cognitive presence is pivotal. This can be achieved by promoting student collaboration, cultivating a climate of empathy, respect, and cooperation, facilitating real-time interactive courses, and maintaining open lines of communication. However, graduate educators should exercise caution when organizing "conferences" within a course, ensuring they align with students' needs and goals.

For Teaching presence, students appreciate a more subdued approach, allowing other facets of the classroom experience to shine. The transition to Zoom-based lectures during the COVID-19 pandemic was met with less enthusiasm and the emphasis was shifted away from traditional course materials, reflecting students' reduced interest in focusing on prescribed content. Consequently, undergraduate educators could focus on refining their teaching skills by incorporating interactive elements like live Q&A sessions or breakout room activities, which encourage student participation. On the other hand, graduate instructors might find that creating easily shareable, multimedia-rich course resources, such as video lectures or interactive simulations, resonates better with students. Moreover, it's important to adjust the intensity of teaching presence to allow other, more motivating facets of the classroom to flourish, as students have demonstrated a reduced inclination to focus on this aspect in the post-COVID classroom environment.

### 5.2. Limitation and further work

The study has certain limitations that should be acknowledged. Firstly, the selection of CoI terms relied on manual interpretation, which could introduce potential biases based on the annotators' background and knowledge. As a result, the finalized library might overlook words that could also represent elements in the CoI framework. To address this limitation, future studies could involve multiple annotators working collaboratively to collect a more comprehensive list of CoI-related terms.

The second limitation is associated with the sentiment analysis. In this study, we conducted sentiment analysis based on the sentence level. However, a student's sentiment may vary when expressing conflicting opinions on different topics within a single sentence. To capture the polarity towards specific topics within their sentence context, a more nuanced sentiment model may be necessary.

Lastly, it is important to note that this study was limited to a single university and focused solely on engineering students. To generalize the findings, further research should encompass students from various academic programs offered at different universities. Additionally, the survey used in this study could be expanded to include a global item that measures overall student evaluations with online learning under emergency circumstance.

## 6. Conclusions

The emergent COVID-19 pandemic presented substantial challenges for educators and students, necessitating a shift towards instructional techniques and learning environments. To investigate the impacts and students' expectations, we conducted a three-year survey study situated within the Project Management program at the University of Maryland.

First, our analysis reveals that there were no significant differences between the two study periods. Despite the challenges posed by the shift, our results indicate that students' performance and self-evaluation remained consistent and comparable to pre-pandemic levels. Second, the study shows significant distinctions in students' expectations, as observed through the CoI lens. Social presence in education is crucial for both undergraduates and graduates. For undergraduates, it connects academic learning with practical job prospects. Graduates value it for networking and mutual support in their professional world. Both seek flexibility in schedules and prefer active learning like projects. Cognitive presence intertwines with social presence. Students exhibit a notable preference for formative exercises, such as projects and assignments, emphasizing their preference for "learning by doing" rather than passive study of course materials. Teaching presence should be more subtle, with students less enthusiastic about Zoom-based lectures.

In summary, these findings underscore the necessity for a significant shift in the traditional classroom model that align with students' expectation and preference. These findings hold meaningful implications for educators who aim to create effective and engaging learning environments, as well as emphasizing the importance of ongoing evaluation and adjustment of educational practices to ensure a high-quality learning experience during the pandemic.